\documentclass[journal]{IEEEtran}%
\usepackage{ifpdf}
\usepackage{graphicx}
\usepackage{amssymb}
\usepackage{amsmath}
\usepackage{cite}
\usepackage{epstopdf}
\usepackage{dsfont}
\usepackage[belowskip=-15pt,aboveskip=0pt,small]{caption}
\usepackage{color}
\usepackage{algorithm}
\usepackage{algorithmicx, algpseudocode}
\usepackage{etoolbox}
\usepackage{subcaption}
\usepackage{balance}

\newcommand{\mbf}[1]{\mathbf{#1}}

\newcommand{\ub}[1]{\underbrace{#1}}

\makeatletter
\def\ScaleIfNeeded{%
\ifdim\Gin@nat@width>\linewidth \linewidth \else \Gin@nat@width \fi
} \makeatother

\usepackage{mathtools}
\DeclarePairedDelimiterX{\inp}[2]{\langle}{\rangle}{#1, #2}
\usepackage{amsmath}

\begin{document}

\title{Energy-Efficient Resource Allocation for
IRS-Assisted Multi-Antenna Uplink Systems}

\author{\IEEEauthorblockN{Ming Zeng, E. Bedeer Mohamed, Octavia A. Dobre, \emph{Fellow, IEEE}, Paul Fortier, and Quoc-Viet Pham
}
\thanks{
M. Zeng and P. Fortier are with Laval University, Quebec, G1V 0A6, Canada (e-mail: mzeng@mun.ca, paul.fortier@gel.ulaval.ca).

E. Bedeer Mohamed is with University of Saskatchewan, Saskatoon, Canada S7N5A9. Email: e.bedeer@usask.ca.

O. A. Dobre is with Memorial University, St. John's, NL A1B 3X9, Canada (e-mail: odobre@mun.ca).

Q. V. Pham is with Pusan National University, Busan, 46241 Korea (e-mail: vietpq@pusan.ac.kr)

} 
}


\maketitle

\begin{abstract}
In this paper, we study the resource allocation for an intelligent reflecting surface (IRS)-assisted uplink system, where the base station is equipped with multiple antennas. 
We propose to jointly optimize the transmit power of the users, active beamforming at the base station, and passive beamforming at the IRS to maximize the overall system energy efficiency while maintaining users' minimum rate constraints.
This problem belongs to a highly intractable non-convex bi-quadratic programming problem, for which an iterative solution based on block coordinate descent is proposed. Extensive simulations are conducted to demonstrate the effectiveness of the proposed scheme and the benefits of having more elements at the IRS.    

\end{abstract}

\IEEEpeerreviewmaketitle


\section{Introduction}
Intelligent reflecting surface (IRS) is a promising technology to boost the network performance for beyond 5G systems \cite{Basar_Access19, Gong_Survey20, Zhao_20}. An IRS is an artificial passive radio planar array, which is composed of multiple low-cost reflecting elements. Each element can add a phase shift to the incident signal, forming various beam patterns. This way, the IRS is capable of tuning the wireless environment to support the data transmission. 

So far, IRS has been applied to a variety of wireless networks \cite{Guo_Arxiv19, Nadeem_Arxiv19, Wu_TWC19, Huang_TWC19, Ding_CL20, Fu_GC19, Yang_Arxiv20, Chen_Access19, Shen_COMML19, Xu_Globecom19, Jamali_Arxiv19}. For example, the IRS-aided multi-user multiple-input single-output (MISO) system has been considered in \cite{Guo_Arxiv19, Nadeem_Arxiv19, Wu_TWC19, Huang_TWC19}, requiring to optimize the active beamforming at the base station (BS) and passive beamforming at the IRS.
More exactly, the authors in \cite{Guo_Arxiv19} consider the weighted sum rate maximization. Lagrangian dual transform is first employed to decouple the original problem. On this basis, the active and passive beamforming is optimized alternatively. The authors in \cite{Nadeem_Arxiv19} aim to maximize the minimum rate among users. The passive beamforming at the IRS is optimized when the channel between the BS and IRS is of rank-one and of full-rank, respectively. The authors in \cite{Wu_TWC19} study the total transmit power minimization subject to users' individual minimum rate constraints. The beamforming for the single-user case is first addressed using alternating optimization, and on this basis, the multi-user case is addressed. The authors in \cite{Huang_TWC19} investigate the system energy efficiency (EE) maximization, and propose two computationally efficient beamforming solutions based on alternating maximization, gradient descent search, and sequential fractional programming. 
To further improve the performance, the IRS-aided MISO non-orthogonal multiple access (NOMA) system is studied in \cite{Ding_CL20, Fu_GC19, Yang_Arxiv20}. 
More specifically, \cite{Ding_CL20} proposes an IRS-NOMA transmission scheme, which ensures that more users are served on each orthogonal spatial direction than spatial division multiple access. The authors in
\cite{ Fu_GC19} and \cite{Yang_Arxiv20} investigate the joint optimization of the transmit beamforming at the BS and the phase shift matrix at the IRS, aiming to minimize the transmit power and maximize the minimum rate, respectively. 
Additionally, IRS is used to enhance physical layer security \cite{Chen_Access19, Shen_COMML19, Xu_Globecom19} and to support millimeter wave transmission \cite{Jamali_Arxiv19}. 

Note that all the above works \cite{Guo_Arxiv19, Nadeem_Arxiv19, Wu_TWC19, Huang_TWC19, Ding_CL20, Fu_GC19, Yang_Arxiv20, Chen_Access19, Shen_COMML19, Xu_Globecom19, Jamali_Arxiv19} consider the downlink transmission. 
In this paper, we consider an IRS-assisted multi-antenna uplink system. To the best of our knowledge, \cite{Zeng_COMML20} is the only work on IRS-based uplink system. However, \cite{Zeng_COMML20} considers the simple scenario when the BS is equipped with a single antenna. Our goal is to maximize the system EE under users' individual minimum rate requirements, involving a joint optimization of power control and beamforming design at both the BS and the IRS. An iterative solution based on block coordinate descent (BCD) is proposed to address the formulated highly intractable non-convex problem. Presented numerical results show the superiority of the proposed scheme over baselines that only optimize some of the three variables. Moreover, it is observed that having more elements at the IRS can yield better system performance improvement than having more antennas at the BS.

\section{System Model and Problem Formulation}
\subsection{System Model}
As shown in Fig.~1, we consider an uplink system, where $K$ single antenna users  communicate with a multi-antenna BS. The number of antennas at the BS is $M$. It is assumed that no direct link exists between the users and BS due to unfavorable propagation conditions. Therefore, this communication takes place via an IRS with $N$ reflecting elements deployed on the facade of a building located in the proximity of both communication ends.


The signal received at the BS is given by 
\begin{align} \label{y}
\mbf{y} =  \sum_{k=1}^K  \mbf{G} \mbf{\Phi} \mbf{h}_{k} \sqrt{P_k} s_k + \mbf{n},
\end{align}
where $s_k$ denotes the signal from user $k$ and is of unit power, i.e., $\mathbb{E}[|s_{k}|^2]=1$, 
$k \in \{1, \cdots, K\}$,
with $\mathbb{E}$ being the expectation operation. 
$P_k$ is the corresponding transmit power of user $k$, satisfying $P_k \leq P_k^{\max}$, with $P_k^{\max}$ being the maximum transmit power. $\mbf{h}_{k} \in \mathbb{C}^{N \times 1}$ denotes the channel vector between user $k$ and IRS, while $\mbf{G} \in \mathbb{C}^{M \times N}$ represents the channel matrix between the IRS and BS. 
%
$\mbf{\Phi} = \rm{diag}(\phi_1, \cdots, \phi_N)$ is an $N \times N$ diagonal matrix whose entries are the $N$ elements $\phi_1, \cdots, \phi_N$,  with the $i$-th diagonal element $\phi_i$ satisfying $|\phi_i| = 1$, $\forall i \in \{1, \cdots, N\}$. The diagonal matrix $\mbf{\Phi}$ accounts for the effective phase shifts from all IRS reflecting elements. The constant modulus constraint $|\phi_i| = 1$, $\forall i \in \{1, \cdots, N\}$, is because the IRS simply reflects the received signal and cannot amplify it.
$\mbf{n} \sim \mathcal{CN} (0, N_0 \mbf{I}_N)$ denotes the additive white Gaussian noise vector at the BS with power spectral density of $N_0$.

\begin{figure}
\centering
\includegraphics[width=0.42\textwidth]{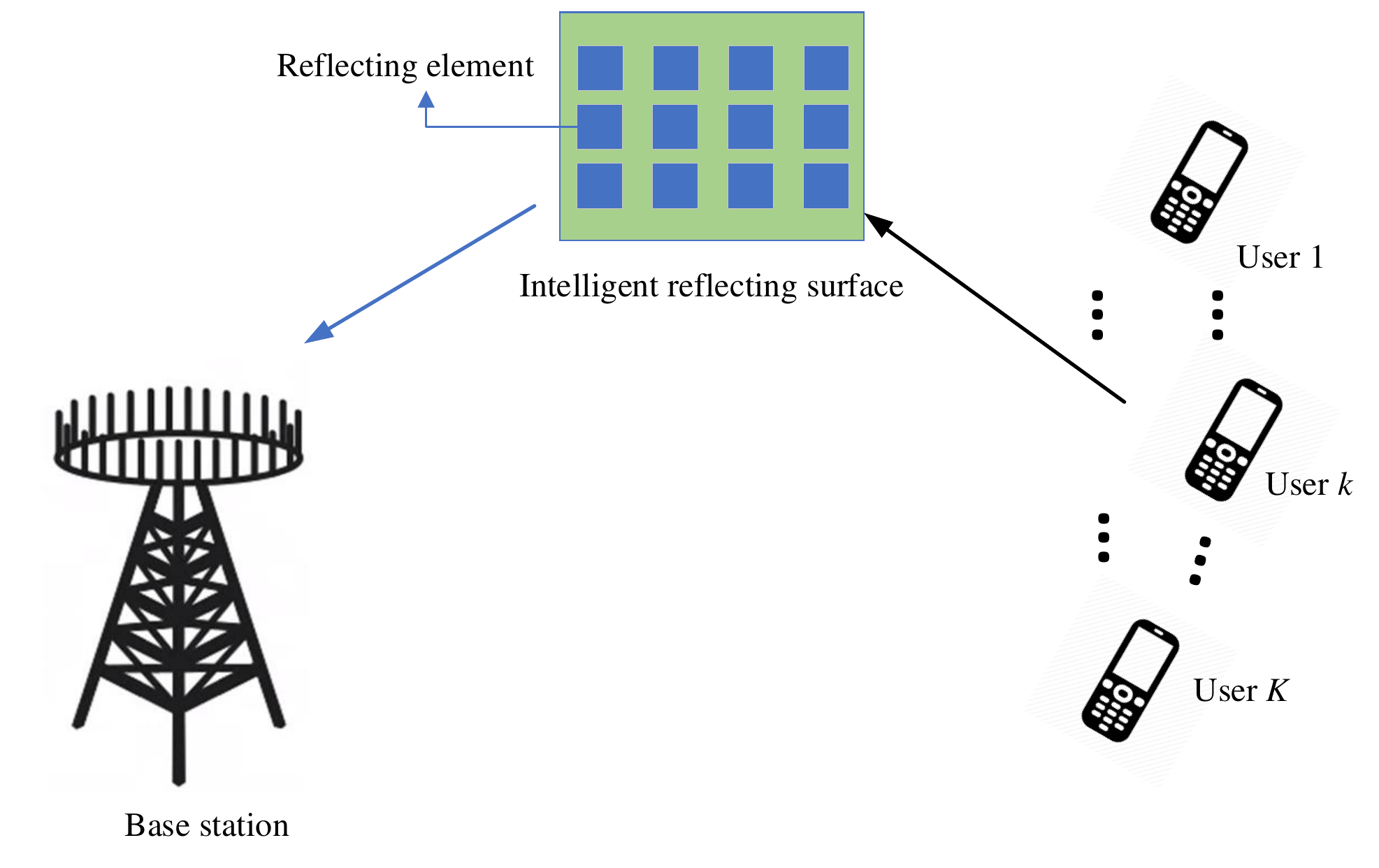}
\caption{IRS-assisted multi-antenna uplink system.}
\end{figure}

By applying the beamforming vector $\mbf{v}_k$ at the BS, we obtain
\begin{align} \label{y}
\mbf{v}_k^H  \mbf{y} =& \mbf{v}_k^H \sum_{i=1}^K  \mbf{G} \mbf{\Phi} \mbf{h}_{i} \sqrt{P_i} s_i + \mbf{v}_k^H \mbf{n} \\ \nonumber
=& \mbf{v}_k^H  \mbf{G} \mbf{\Phi} \mbf{h}_{k} \sqrt{P_k} s_k +   \sum_{i\neq k}^K  \mbf{v}_k^H \mbf{G} \mbf{\Phi} \mbf{h}_{i} \sqrt{P_i} s_i + \mbf{v}_k^H \mbf{n} .
\end{align}
Accordingly, the signal-to-interference-plus-noise (SINR) of user $k$ can be expressed as
\begin{equation} \label{SINR}
\gamma_k = \frac{|\mbf{v}_k^H  \mbf{G} \mbf{\Phi} \mbf{h}_{k}|^2 P_k}{\sum_{i \neq k} |\mbf{v}_k^H \mbf{G} \mbf{\Phi} \mbf{h}_{i}|^2 P_i + |\mbf{v}_k^H \mbf{n}|^2},
\end{equation}
and the achievable data rate of user $k$ is given by $R_k = \log_2 (1+ \gamma_k)$.

\subsection{Problem Formulation}
In this paper, we aim to maximize the system EE under users' quality-of-service (QoS) constraints.  
The considered problem can be formulated as follows:
\begin{subequations}\label{P1}
\begin{align} 
 \underset{\mbf{P}, \mbf{V}, \mbf{\Phi}}{\max} &~ \frac{ \sum_{k=1}^K \log_2 (1+ \gamma_k) }{ \psi \sum_{k=1}^K P_k +P_c } \\
\text{s.t.}
& ~ \gamma_k \geq \gamma_k^{\min}, \quad \forall k \in \{1, \cdots, K\} \\
& ~ P_k \leq P_k^{\max}, \quad  \forall k \in \{1, \cdots, K\} \\
& ~ |\phi_i| = 1, \quad  \forall i \in \{1, \cdots, N\},
\end{align}
\end{subequations}
where $\mbf{P} = [P_1, \cdots, P_K]$ and $\mbf{V} = [\mbf{v}_1, \cdots, \mbf{v}_K]$ are the transmit power vector and the receiver beamforming matrix, respectively. $\psi$ is a constant accounting for the  power amplifier inefficiency. $P_{\rm{C}} $ denotes the fixed circuit power consumption of all the users.

\section{Proposed Solution}
One can show that the problem in  \eqref{P1} is non-convex and it requires the joint optimization of the passive beamforming at the IRS, active beamforming at the BS, and power control at the users. In particular, the main difficulty in solving \eqref{P1} lies in the coupling among the optimization variables.  
To address this difficulty, we adopt BCD to decompose the problem into three subproblems of $\mbf{P}$, $\mbf{V}$ and $\mbf{\Phi} $, and solve them alternatively.

\subsection{Power Control at the Users}
Under given $\mbf{V}$ and $ \mbf{\Phi}$, the power control problem can be simplified as 
\begin{subequations}\label{P1_power}
\begin{align} 
 \underset{\mbf{P}}{\max} &~ \frac{ \sum_{k=1}^K \log_2 (1+ \gamma_k) }{ \psi \sum_{k=1}^K P_k +P_c } \\
\text{s.t.}
& ~ \gamma_k \geq \gamma_k^{\min}, \forall k \in \{1, \cdots, K\} \\
& ~ P_k \leq P_k^{\max}, \forall k \in \{1, \cdots, K\}.
\end{align}
\end{subequations}
It is clear that (\ref{P1_power}) is a fractional problem, which can be transformed into a series of parametric subtractive-form subproblems as follows:
\begin{subequations} \label{eq11}
\begin{align} 
\underset{\bf{P}} {\rm{max}}~ &{\sum_{k=1}^K \log_2 (1+ \gamma_k) }\!-\!\lambda^{(l-1)}\left({\psi\sum_{k=1}^K P_k \!+\!P_{\rm{C}}}\right) \\
{\rm{s.t.}} &\;(\rm{\ref{P1_power}b}),(\rm{\ref{P1_power}c}),
\end{align}
\end{subequations}
where $l$ is the iteration index and $\lambda^{(l-1)}$ is a non-negative parameter. Starting from $\lambda^{(0)}=0$, $\lambda^{(l)}$ can be updated as $\lambda^{(l)} = \frac{\sum_{k=1}^K \log_2 (1+ \gamma_k^{(l)}) }{\psi\sum_{k=1}^K P_{k}^{(l)}+P_{\rm{C}}}$, {\color{black}{where $\gamma_k^{(l)}$ and $P_{k}^{(l)}$ are the updated SINRs and power}} after solving (\ref{eq11}). Moreover, the maximum value of (\ref{eq11}) is calculated as $\varepsilon^{(k)}\! = \!\sum_{k=1}^K  \log_2 (1+ \gamma_k^{(l)})  \!-\!\lambda^{(l-1)} (\psi\sum_{k=1}^K P_{k}^{(l)}+P_{\rm{C}})$. As shown in \cite{W_dink}, $\lambda^{(l)}$ keeps growing while $\varepsilon^{(l)}$ keeps declining as $l$ increases. When $\varepsilon^{(l)}=0$, $\lambda^{(l)}$ is maximized, which is also the maximum EE of (\ref{P1_power}).

The problem now lies in how to solve (\ref{eq11}) for a given $\lambda$. It is clear that (\ref{eq11}) is a non-convex optimization problem due to the non-concave objective function. Let us denote $g_{k,i} = {|\mbf{v}_k^H \mbf{G} \mbf{\Phi} \mbf{h}_{i}|^2 }/{ |\mbf{v}_k^H \mbf{n}|^2} $ to simplify the used notation.
After a straightforward mathematical manipulation, the objective function in (\ref{eq11}) can be re-written as
\begin{align}
f=&\ub{\sum_{k=1}^K \log_2 \left( \sum_{i =1}^K P_i g_{k,i}+ 1  \right)   - \lambda  \bigg( \psi \sum_{k=1}^K P_k   + P_c \bigg)}_{f_1(\mathbf{P})}\nonumber \\
&-\ub{\sum_{k=1}^K \log_2 \left( \sum_{i \neq k}^K P_i g_{k,i}+ 1  \right) }_{f_2(\mathbf{P})},
\end{align}
and the QoS constraints in (\rm{\ref{eq11}b}) can be reformulated as
\begin{equation} \label{QoS}
{P_k g_{k,k}}\geq \gamma_k^{\min} \left({ \sum_{i \neq k}^K P_i g_{k,i}+ 1}\right), \forall k \in \{1, \cdots, K\},
\end{equation}
which is an affine constraint. Then, the power control optimization problem can be rewritten as
\begin{equation} \label{dc_programming}
\underset{\bf{P}}{\rm{max}}~~ f_1(\mathbf{P})-f_2(\mathbf{P}), ~{\rm{s.t.}}~   \eqref{QoS}, (\rm{\ref{P1_power}c}),
\end{equation}
where both functions $ f_1(\mathbf{P})$ and $f_2(\mathbf{P})$ are concave functions. Thus, the objective $f_1(\mathbf{P})-f_2(\mathbf{P})$ is a DC function (difference of two concave functions). For $k \in \{1, \cdots, K\}$, define the vector $\mathbf{e}_k \in \mathbb{R}^{K}$, satisfying $\mathbf{e}_k(k) = 0$ and $\mathbf{e}_k(i) = \frac{g_{k,i}}{\ln2}, ~ i \neq k$. The gradient of $f_2$ at $\mathbf{P}$ is given by
\begin{equation} \label{power_gradient}
\nabla f_2({\mathbf{P}})= \sum_{k=1}^K \frac{\mathbf{e}_k}{1+ \sum_{i\neq k}g_{k,i}P_{i}}.
\end{equation} 
The following procedure generates a sequence $\{{\bf{P}}^{(l)} \} $ of improved feasible solutions. Initialized from a feasible $\{{\bf{P}}^{(0)} \} $, $\{{\bf{P}}^{(l)} \} $ is obtained as the optimal
solution of the following convex problem at the $l$-th iteration:
\begin{align} \label{dc_convex}
\underset{\bf{P}} {\rm{max}} ~~ &f_1({\bf{P}})-f_2({\bf{P}}^{(l-1)})- \inp{\nabla f_2({\bf{P}}^{(l-1)})}{{\bf{P}} -{\bf{P}}^{(l-1)}} ~  \nonumber \\
{\rm{s.t.}} &\; \eqref{QoS}, (\rm{\ref{P1_power}c}),
\end{align}
{\color{black}where $ \inp{\cdot}{\cdot}$ denotes the inner product operation.} 
Note that \eqref{dc_convex} can be efficiently solved by available convex software packages \cite{Cvx}.

\subsection{Active Beamforming at the BS}
Under given $ \mbf{\Phi}$ and $ \mbf{P}$, it can be seen that the SINR of user $k$ only depends on $\mbf{v}_k$ according to \eqref{SINR}. For notational simplicity, denote $\bar{\mbf{h}}_{i} =\mbf{G} \mbf{\Phi} \mbf{h}_{i}$. 
The optimal beamforming vector which can balance the interference and noise is
given by
\begin{equation}\label{eq:v}
\mbf{v}_k = \left( N_0 \mbf{I}_N + \sum_{ i \neq k} P_i \bar{\mbf{h}}_{i} \bar{\mbf{h}}_{i}^H \right)^{-1} \bar{\mbf{h}}_{k},
\end{equation}
and is referred to as the minimum mean square error (MMSE) receiver \cite{4}. The corresponding output SINR in this case is given as $\gamma_k =P_k \bar{\mbf{h}}_{k}^H \left( N_0 \mbf{I}_N + \sum_{ i \neq k} P_i \bar{\mbf{h}}_{i} \bar{\mbf{h}}_{i}^H \right)^{-1} \bar{\mbf{h}}_{k}.$

\subsection{Passive Beamforming at the IRS}
Under given $\mbf{V}$ and $ \mbf{P}$, the beamforming optimization at the IRS can be simplified as
\begin{subequations}\label{P1_IRS}
\begin{align} 
 \underset{\mbf{\Phi}}{\max} &~ \sum_{k=1}^K \log_2 (1+ \gamma_k)  \\
\text{s.t.}
& ~ \gamma_k \geq \gamma_k^{\min}, \forall k \in \{1, \cdots, K\} \\
& ~ |\phi_i| = 1, \forall i \in \{1, \cdots, N\}.
\end{align}
\end{subequations}
To easily handle the term $|\mbf{v}_k^H \mbf{G} \mbf{\Phi} \mbf{h}_{i}|^2$ that appears in the SINR expression, we first re-arrange the diagonal matrix $\mbf{\Phi}$ into a vector $\mbf{w} \in \mathbb{C}^{N \times 1}$, with element $w_j = \phi_j^H$, $\forall j \in \{1, \cdots, N\}$. It is clear that $\mbf{w}$ contains all the information of $\mbf{\Phi}$. Then, we introduce an auxiliary vector $\widehat{\mbf{h}}_{k,i} = ( \mbf{v}_k^H\mbf{G} ) \circ \mbf{h}_{i}$, where the operation $\circ $ represents the Hadamard product. Accordingly, it can be easily verified that the following equality holds $| \mbf{v}_k^H \mbf{G} \mbf{\Phi} \mbf{h}_{i}|^2= |\mbf{w}^H \widehat{\mbf{h}}_{k,i}|^2$.
Then, problem \eqref{P1_IRS} can be re-expressed as
\begin{subequations}\label{P3}
\begin{align} 
 \underset{\mbf{w}}{\max} &~   \sum_{k=1}^K \log_2 \left(1+ \frac{|\mbf{w}^H \widehat{\mbf{h}}_{k,k}|^2 P_k}{ \sum_{i \neq k}^K  |\mbf{w}^H \widehat{\mbf{h}}_{k,i}|^2 P_i + \sigma_k^2  }  \right)  \\
\text{s.t.}
& ~ |\mbf{w}^H \widehat{\mbf{h}}_{k,k}|^2 P_k \geq \gamma_k^{\min} \left( \sum_{i \neq k}^K  |\mbf{w}^H \widehat{\mbf{h}}_{k,i}|^2 P_i + \sigma_k^2 \right), \nonumber \\ 
& ~ |w_i| = 1, \qquad \forall i \in \{1, \cdots, N\},
\end{align}
\end{subequations}
where $\sigma_k^2 =|\mbf{v}_k^H \mbf{n}|^2$.
To address \eqref{P3}, we further reformulate $ |\mbf{w}^H \widehat{\mbf{h}}_{k,i}|^2 P_i$ as
\begin{align}
&|\mbf{w}^H \widehat{\mbf{h}}_{k,i}|^2 P_i  =   \mbf{w}^H  P_i \widehat{\mbf{h}}_{k,i} \widehat{\mbf{h}}_{k,i}^H \mbf{w} \nonumber  \\ 
&= {\rm{Tr}} ( \mbf{w}^H  P_i \widehat{\mbf{h}}_{k,i} \widehat{\mbf{h}}_{k,i}^H \mbf{w} ) = {\rm{Tr}} (   P_i \widehat{\mbf{h}}_{k,i} \widehat{\mbf{h}}_{k,i}^H \mbf{w} \mbf{w}^H ).
\end{align}
Now, we introduce two auxiliary matrices $\mbf{H}_{k,i}= P_i  \widehat{\mbf{h}}_{k,i} \widehat{\mbf{h}}_{k,i}^H$ and $\mbf{W}= \mbf{w} \mbf{w}^H $. It can be easily verified that $\mbf{H}_{k,i}$ and  $\mbf{W}$ are positive semi-definite matrices. Meanwhile, problem \eqref{P3} can be re-expressed as
\begin{subequations} \label{P5}
\begin{align} 
 \underset{\mbf{W}}{\max} &~   \sum_{k=1}^K \log_2 \left(1+ \frac{{\rm{Tr}} (\mbf{W} \mbf{H}_{k,k}) }{ \sum_{i \neq k}^K  {\rm{Tr}} (\mbf{W} \mbf{H}_{k,i}) + \sigma_k^2  }  \right)  \\
\text{s.t.}
& ~ {\rm{Tr}} (\mbf{W} \mbf{H}_{k,k}) \geq \gamma_k^{\min} \left( \sum_{i \neq k}^K   {\rm{Tr}} (\mbf{W} \mbf{H}_{k,i})  + \sigma_k^2 \right), \\
& ~ {\rm{diag}} \{ \mbf{W} \} = 1,  \\
& ~ \mbf{W} \succeq \mbf{0}, \\
& ~ {\rm{rank}} (\mbf{W})=1,
\end{align}
\end{subequations}
where $\rm{diag} \{\mbf{W}\} $ returns the diagonal elements of $\mbf{W}$. Thus, ${\rm{diag}} \{ \mbf{W} \} = 1$ is equivalent to $ w_i^2=1$, and further $|w_i| = 1$, $\forall i \in \{1, \cdots, N\}$. 

Problem \eqref{P5} is still non-convex due to the non-convex objective function (\ref{P5}a) and the rank one constraint (\ref{P5}e). The 
rank one constraint (\ref{P5}e)  can be replaced with a convex positive semidefinite constraint $\mbf{W} -  \bar{\mbf{w}}\bar{\mbf{w}}^H   \succeq \mbf{0}$, 
where $\bar{\mbf{w}} \in \mathbb{C}^{N \times 1} $ is an auxiliary variable. 
Further,  $\mbf{W} -  \bar{\mbf{w}}\bar{\mbf{w}}^H   \succeq \mbf{0}$ can be replaced with its convex Schur complement as follows:
\begin{equation} \label{rank-1 Schur} 
   \begin{bmatrix}
    \mbf{W}  &  \bar{\mbf{w}}\\
    \bar{\mbf{w}}^H &  1
  \end{bmatrix} \succeq \mbf{0}.
\end{equation}
Now, let us study the objective function (\ref{P5}a), which can be re-expressed as \eqref{objective_diff} at the top of next page.
\begin{figure*}[!t]
\begin{align} \label{objective_diff}
 \sum_{k=1}^K \log_2 \left(1+ \frac{{\rm{Tr}} (\mbf{W} \mbf{H}_{k,k}) }{ \sum_{i \neq k}^K  {\rm{Tr}} (\mbf{W} \mbf{H}_{k,i}) + \sigma_k^2  }  \right)  
=&{\sum_{k=1}^K \log_2 \left( { \sum_{i =1}^K  {\rm{Tr}} (\mbf{W} \mbf{H}_{k,i}) + \sigma_k^2  }  \right)}
-{\sum_{k=1}^K  \log_2 \left( { \sum_{i \neq k}^K  {\rm{Tr}} (\mbf{W} \mbf{H}_{k,i}) + \sigma_k^2  }  \right)} \\ \nonumber
=&\ub{\sum_{k=1}^K \log_2 \left( {  {\rm{Tr}} (\mbf{W} \sum_{i =1}^K  \mbf{H}_{k,i}) + \sigma_k^2  }  \right)}_{f_3(\mathbf{W})} 
-\ub{\sum_{k=1}^K  \log_2 \left( {   {\rm{Tr}} (\mbf{W} \sum_{i \neq k}^K \mbf{H}_{k,i}) + \sigma_k^2  }  \right)}_{f_4(\mathbf{W})}.
\end{align}
\hrulefill
\end{figure*} 
\begin{figure*}[!t]
\begin{subequations} \label{P6}
\begin{align} 
 \underset{\mbf{W}, \bar{\mbf{w}} }{\max} &~  f_3 (\mathbf{W}) -f_4({\bf{W}}^{(l-1)})- \sum_{n=1}^N \sum_{j=1}^{n-1} \mathfrak{R} \left(  \mbf{W}(n,j)  - \mbf{W}(n,j)^{(l-1)} \right) \times  \frac{\partial f_4}{\partial \mathfrak{R} ( \mbf{W}(n,j)^{(l-1)})} \\ 
 &~- \sum_{n=1}^N \sum_{j=1}^{n-1}  \mathfrak{I} \left(  \mbf{W}(n,j)  - \mbf{W}(n,j)^{(l-1)} \right) \times  \frac{\partial f_4}{\partial \mathfrak{I} ( \mbf{W}(n,j)^{(l-1)})} \nonumber \\
\text{s.t.}
& ~ (\ref{P5}\rm{b})-(\ref{P5}\rm{d}), \eqref{rank-1 Schur}.
\end{align}
\end{subequations}
\hrulefill
\end{figure*}
It can be seen that \eqref{objective_diff} is a DC function, and thus, can be handled using DC programming. However, we cannot simply use the partial derivative for real variables as in \eqref{power_gradient} since $\mathbf{W}$ is a complex matrix. Moreover, the real function $f_4(\mathbf{W})$ is not complex-analytic (holomorphic), and thus, its derivative with respect to $\mathbf{W} $ does not exist in the conventional sense of a complex derivative \cite{Delgado09}.
To address this issue, we need to calculate the derivatives of $f_4$ over both the real and imaginary parts of $\mathbf{W} $. {\color{black}According to the symmetry of $\mathbf{W} $ and ${\rm{diag}} \{ \mbf{W} \} = 1$, we only need to calculate the real and imaginary parts of the elements below the diagonal of $\mathbf{W}$.} For $\mbf{W}(n,j), \forall n \in \{1, \dots, N\}, j <n$, we have 
\begin{align*}
\frac{\partial f_4}{\partial \mathfrak{R} (\mbf{W}(n,j ) )}  &=
\frac{1}{\ln 2}  \sum_{k=1}^K \frac{ \sum_{i \neq k}^K \mbf{H}_{k,i} (n,j) + \mbf{H}_{k,i}^H (n,j) }{   {\rm{Tr}} (\mbf{W} \sum_{i \neq k}^K \mbf{H}_{k,i}) + \sigma_k^2}, 
\end{align*}
\begin{align*}
\frac{\partial f_4}{\partial \mathfrak{I} (\mbf{W}(n,j ) )}  &=
\frac{-i}{\ln 2}  \sum_{k=1}^K \frac{ \sum_{i \neq k}^K \mbf{H}_{k,i} (n,j) - \mbf{H}_{k,i}^H (n,j) }{   {\rm{Tr}} (\mbf{W} \sum_{i \neq k}^K \mbf{H}_{k,i}) + \sigma_k^2},
\end{align*}
where $\mathfrak{R}(\cdot)$ and $\mathfrak{I}(\cdot)$ are the operations of obtaining the real and imaginary parts of a complex variable, respectively. Additionally, $\mathbf{W}(n,j)$ and $\mbf{H}_{k,i}(n,j)$ denote the $(n,j)$-th element of $\mathbf{W}$ and $\mbf{H}_{k,i}$, respectively.

Then, at the $l$-th iteration of the DC programming, problem \eqref{P5} can be re-written as \eqref{P6} at the top of next page.
It is clear that \eqref{P6} is a semidefinite programming (SDP), and is convex. Therefore, the optimal solution can be obtained using standard convex optimization methods, such as the interior-point method. 
Now the question becomes how to convert the global solution $\mbf{W}^\star $ and $ \bar{\mbf{w}}^\star $ of \eqref{P6} (after dropping rank one constraint) to a feasible solution of the original problem \eqref{P5} (that has a rank one constraint). We use Gaussian randomization to address this issue, where the solution is considered as the mean of a multivariate Gaussian random vector with $\mbf{W}^\star- \bar{\mbf{w}}^\star \bar{\mbf{w}}^{\star H}$ as its covariance matrix. In particular, we generate $Q$ trials of a random variable $ \mathcal{E}_q  \sim (\mbf{w}^\star, \mbf{W}^\star- \bar{\mbf{w}}^\star \bar{\mbf{w}}^{\star H} ) $, $q = 1, ..., Q$. Since $\mathcal{E}_q$ is not guaranteed to be feasible, we apply a rescaling factor to obtain the feasible candidate solution vector $\mbf{w}_{\mathcal{E}_1}, \cdots, \mbf{w}_{\mathcal{E}_Q}$. Then, we select the value of $\mbf{w}_{\mathcal{E}_q}$ which maximizes the objective. 

\subsection{Iterative Update until Convergence}
The EE is maximized after we repeat the following steps until convergence: 1) solve \eqref{dc_convex} iteratively to find $\mbf{P}$; 2) calculate $\mbf{V}$ using \eqref{eq:v}; and 3) solve \eqref{P6} iteratively and use Gaussian randomization to find $\mbf{\Phi}$.
Note that convergence is guaranteed since the EE increases or remains unchanged at each iteration and the EE clearly has an upper bound.

\section{Simulation Results}
In this section, simulations are conducted to verify the effectiveness of the proposed solution. The default simulation parameters are set as follows \cite{Guo_Arxiv19, Wu_TWC19}: the IRS is composed of 4 elements, while the BS is equipped with 4 antennas. Additionally, the number of users is 3. The distance between the BS and IRS and that between the IRS and users are generated randomly with a uniform distribution {\color{black}within 50 and 100 m, respectively}. The small-scale fading between the BS and IRS and that between the IRS and users are modeled by the Rician (K-factor is 5) and Rayleigh fading, respectively. Meanwhile, the corresponding large-scale path-loss of these two channels follows $30+24 \log_{10}(d)$ and  $30+28 \log_{10}(d)$, respectively, where $d$ is the distance in meters. The bandwidth is $B = 1$ MHz, while the noise power spectral density is $N_0=-174$ dBm/Hz. Additionally, the fixed power consumption $P_c$ is $50$ mW, while the power inefficiency $\psi$ is $0.35$. For simplicity, we assign the same maximum transmit power and minimum rate requirement to all users.

We consider the following baseline schemes: 1) jointly optimize $\mbf{P}$ and $\mbf{V}$ under fixed $\mbf{\Phi}$, referred to as ``Fix IRS''; 2) jointly optimize $\mbf{P}$ and $\mbf{\Phi}$ under fixed $\mbf{V}$, referred to as ``Fix BS''; 3) jointly optimize $\mbf{V}$ and $\mbf{\Phi}$ under fixed $\mbf{P}$, referred to as ``Fix PA''; and 4) randomly set the values for all the variables, referred to as ``Fix all''. 

\begin{figure}
	\centering
	\includegraphics[width=0.5\textwidth]{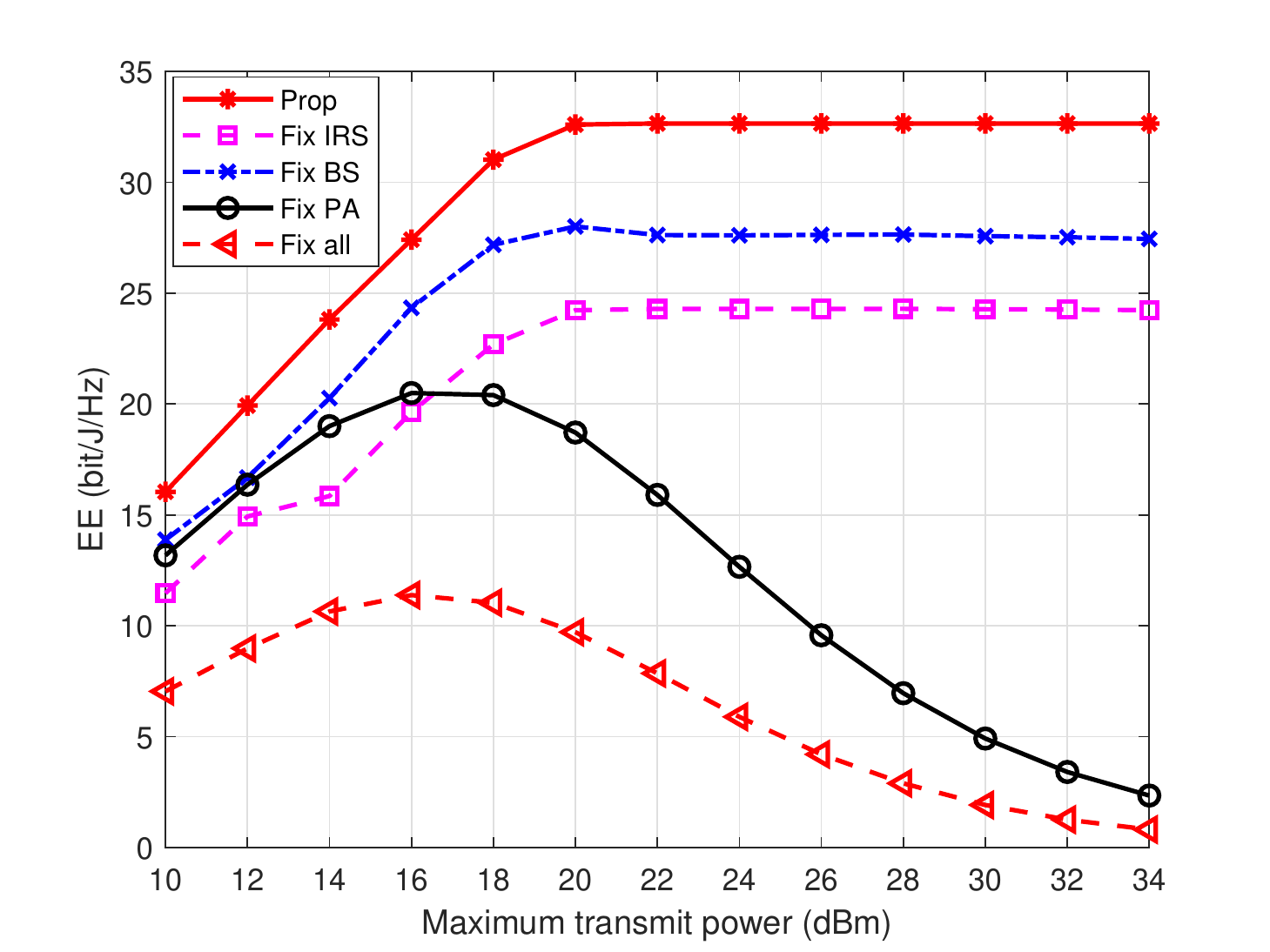}
	\caption{EE versus maximum transmit power for all schemes.}
	\label{SINR}
\end{figure}

Figure \ref{SINR} compares the proposed scheme with the baseline ones as the maximum transmit power increases. Clearly, the proposed scheme, i.e., ``Prop'' is the best, while ``Fix all'' is the worst. This shows the necessity of optimizing all three variables. Meanwhile, the EE for ``Prop'', ``Fix IRS'' and ``Fix BS'' first grows with the maximum transmit power, and then saturates. This can be explained by the concavity of the $\log$ function for the rate, which results in {\color{black}some power unused in order to maximize the EE under high maximum transmit power}. In contrast, when the power is not controlled properly, e.g., in ``Fix PA'' and ``Fix all'', increasing the maximum transmit power may lead to a decrease in EE. 

\begin{figure}
\centering
\includegraphics[width=0.5\textwidth]{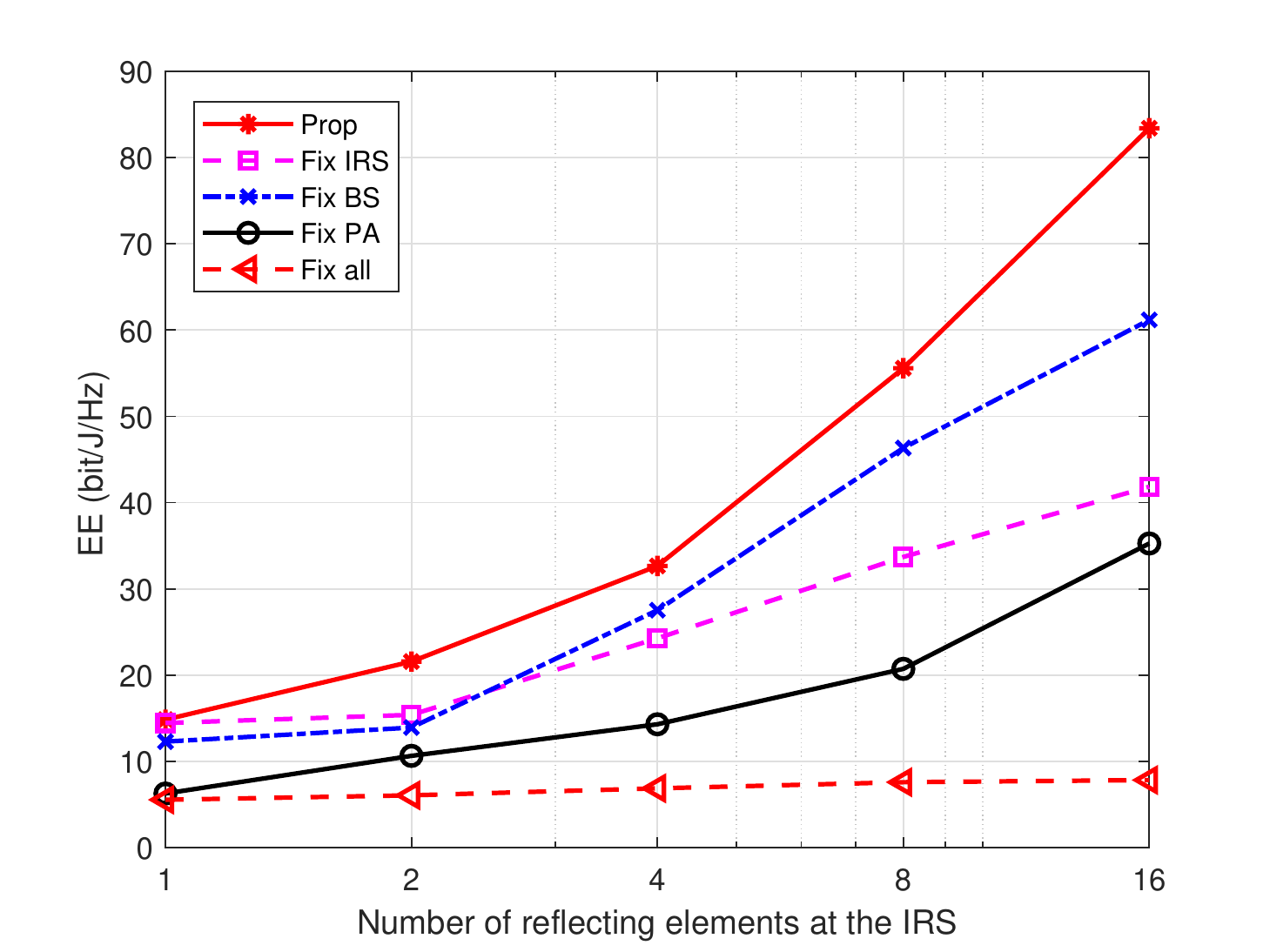}
\caption{EE versus the number of reflecting elements at the IRS for all schemes.}
\label{IRS}
\end{figure}

\begin{figure}
\centering
\includegraphics[width=0.5\textwidth]{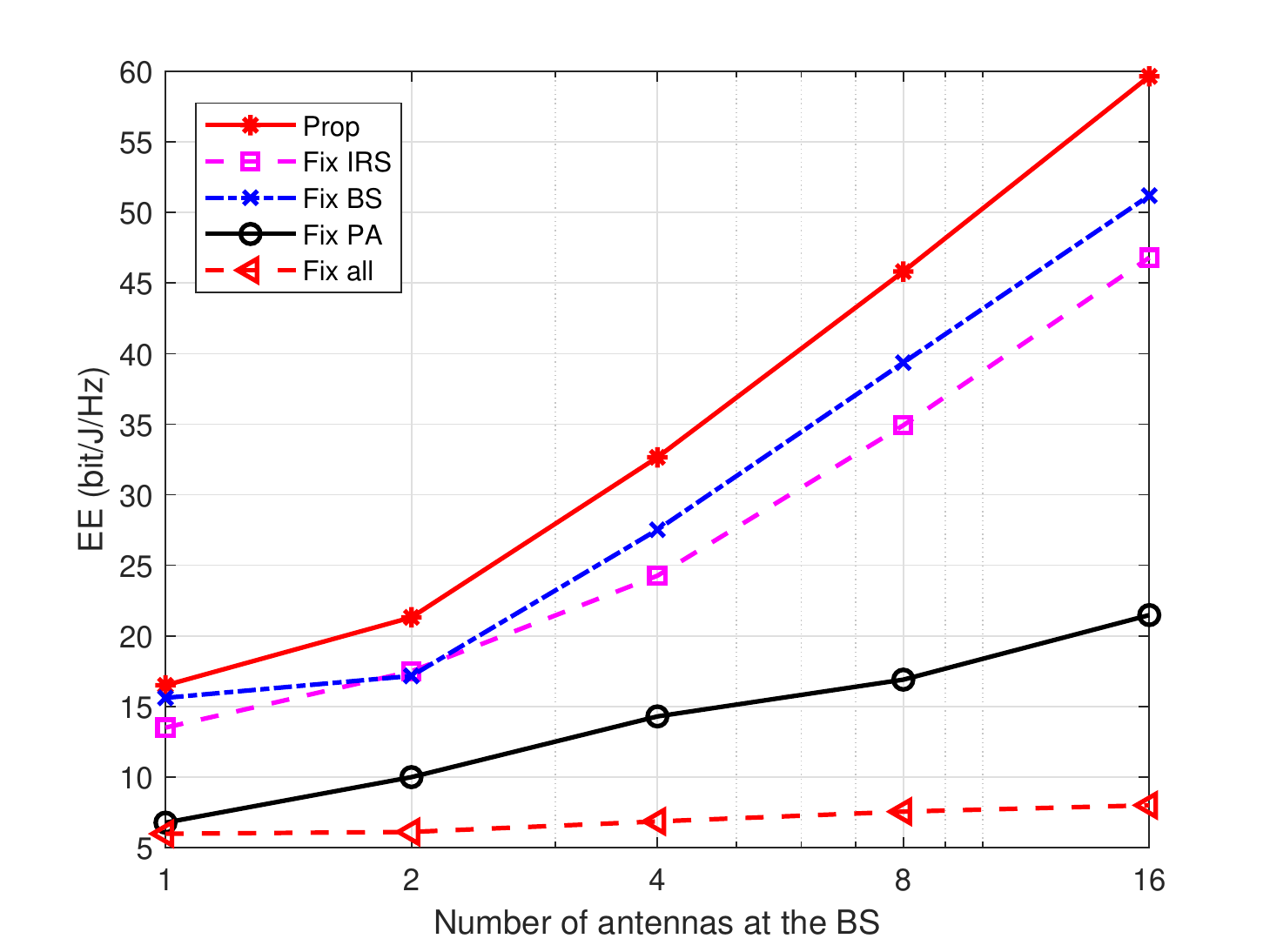}
\caption{EE versus the number of antennas at the BS for all schemes.}
\label{BS}
\end{figure}

Figure \ref{IRS} plots the EE versus the number of reflecting elements at the IRS for all schemes. 
It is clear that the EE for all schemes increases with the number of reflecting elements at the IRS, showing the effectiveness of deploying more elements at the IRS. Among all schemes, the proposed scheme still dominates all baselines, verifying again the need for joint optimization of the power and beamforming at both the IRS and BS. Additionally, Fig. \ref{BS} shows the EE versus the number of antennas at the BS for all schemes. Likewise, the EE grows with the number of antennas at the BS for all schemes. Meanwhile, the proposed scheme still achieves the best performance. By comparing Fig. \ref{IRS} with Fig. \ref{BS}, it can be seen that having more elements at the IRS can lead to higher EE improvement than having more antennas at the BS. This is probably because the IRS elements affect both the channel between the BS and the IRS and that between the IRS and the users. In contrast, the antenna at the BS only affects the channel between the BS and the IRS.

\section{Conclusion}
In this paper, we studied the EE maximization for a multi-user multi-antenna uplink system with the aid of IRS. 
Three different variables, namely the transmit power at the users, phase shift at the IRS and beamforming matrix at the BS need to be jointly optimized. An iterative solution based on block coordinate descent was proposed, which handles one variable at one iteration. Numerical results showed that the proposed scheme dominates the baselines that optimize only some of the variables. Results also indicated that increasing the number of IRS elements leads to better improvements of the system EE when compared to increasing the number of antennas the BS.

\bibliographystyle{IEEEtran}
\balance
\bibliography{IEEEabrv,conf_short,jour_short,mybibfile}

\end{document}